\documentclass[%
reprint,
groupedaddress,
 amsmath,amssymb,
 aps,
prstab,
floatfix,
]{revtex4-1}

\usepackage{graphicx}
\usepackage{dcolumn}
\usepackage{bm}

\begin{document}

\newcommand{\I}{\mathrm{i}}
\newcommand{\E}{\mathrm{e}}
\newcommand{\D}{\,\mathrm{d}}
\newcommand{\s}{\mathrm{sign}}
\newcommand{\sinc}{\mathrm{sinc}}

\title{Statistics of relativistic electrons radiating in periodic fields}

\author{Eugene Bulyak}
\email{bulyak@kipt.kharkov.ua}
\altaffiliation[Also at ]{V.N.~Karazin National University, 4 Svodody sq., Kharkiv, Ukraine}
\affiliation{National Science Center `Kharkov Institute of Physics and Technology', 1 Academichna str, Kharkiv, Ukraine}

\author{Nikolay Shul'ga}
\altaffiliation[Also at ]{V.N.~Karazin National University, 4 Svodody sq., Kharkiv, Ukraine}
\affiliation{National Science Center `Kharkov Institute of Physics and Technology', 1 Academichna str, Kharkiv, Ukraine}

\date{\today}

\begin{abstract}
We developed a general method for assessing the evolution of the energy spectrum of relativistic charged particles that have undergone small quantum losses, such as the ionization losses when the electrons pass through matter and the radiation losses in the periodic fields. These processes are characterized by a small magnitude of the recoil quantum as compared with the particle's initial energy. We convey the statistical consideration of the radiating electrons and demonstrate that at a small average number of the recoils, the electron's spectrum can be described as a composition of consecutive convolutions of the recoil spectrum with itself, taken with the Poisson mass. In this stage, the electron's spectrum reveals some individual characteristics of the recoil spectrum. Furthermore, the spectrum loses individuality and allows for proximate description in the terms of statistical parameters. This consideration reveals that the width of the electron's spectrum is increasing with the number of recoils according to the power law, with the power index being inverse to the stability parameter, which gradually increases  with the number of recoils from one to two. Increase of the spectrum width limits the ability of the beam to generate coherent radiation in hard x-ray and gamma-ray region.
\end{abstract}


\maketitle

\section*{INTRODUCTION}
In a number of processes involving beams of high-energy electrons, such as radiation in the periodic structures, ionization losses in matter, etc., the energy degradation of an incident electron is in the form of the small portions (recoil quanta), which spectrum is almost independent of the electron's energy. In our previous papers \cite{bulyak17b,bulyak18}, we have considered the evolution of the spectrum of such electron beam. It was shown, that the spectrum is determined by the parameters of a single recoil and the average number of the recoils.

This paper is concentrated on the dependence of the spectrum width on the number of recoils in the intermediate range, in between a small average number where the electron's spectrum can be described as a composition of consecutive convolutions of the recoil spectrum with itself taken with the Poisson mass, and the diffusion limit where the width increases as square root of the number of recoils.

This paper is organized as follows: in the first section, we present a method of assessing the evolution of the straggling function that describes the distribution of the energy losses in the interim range of the number of recoils. In the second section, we validate the method by comparing it to the known theories at the limiting cases. The third section presents the results of the study of the kinetics of the radiating electrons in short undulators. The fourth section summarizes the results.

\section{STATISTICS OF THE RADIATING ELECTRONS}

\subsection{Preliminaries}
Distinguishing features of the considered system are: (i) Big number of the ensemble members  (electrons in the radiating bunch) $n\sim 10^{10}$. (ii) Small average number of recoils (defined as ratio of the energy emitted by the electron to the mean energy of the spectrum of the radiation) $x < 10^4 $.

We adopt the assumption, that the spectrum of emitting quantum of the radiation inducing energy loss (recoil), $w(\omega)$ is `physical': it has compact support, $0\le\omega_\text{min}\le \omega \le\omega_\text{max}<\infty$ with $\omega $ being the energy of the recoil.
The spectrum is normalized to unity, $\int w(\omega)\D \omega =1$. (Here and below we drop out the infinite limits in integration.)

In this paper, we use the reduced energy units: $\epsilon$ for the energy in the straggling spectrum, $\omega$ for the energy of the spectrum of the recoil quantum, both are dimensionless, normalized to the energy unit, e.g., to the charged particle rest energy, \cite{bulyak18}. We use a convention for the Fourier transform in the form of:
\[
(\mathcal{F}f)(s)=\hat{f}(s) = \int\E^{-2\I\pi \omega s} f(\omega ) \D\omega
\]
with $s$ being the variable in the Fourier transform domain that complements to $\omega $ (or $\epsilon $). For the inverse Fourier transform, $(\mathcal{F}^{-1}f)(s)=\check{f}(s)$,  the $(-)$ sign in the exponent of the integrand is replaced with $(+)$ sign.

\begin{figure} 
   \includegraphics[width=0.8\columnwidth]{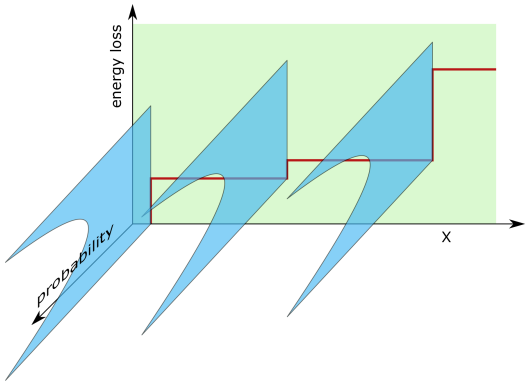}
   \caption{A sketch of straggling process. The distribution of recoil magnitudes resembles the dipole radiation spectrum.}
   \label{fig:scheme}
\end{figure}

A sketch of the straggling electron's trajectory in the plane $(x,\epsilon)$ is presented in Fig.~\ref{fig:scheme}. The trajectory is composed of the free paths of random length, with the mean unit  value and the (positive) random jumps having the same probability density distribution $w(\omega)$. Such a process belongs to the subclass of the \emph{Subordinate to the Compound Poisson Process,}  which in turn belongs to the $\alpha$-stable (or L\'{e}vy) processes, see, e.g., \cite{applebaum04}.

The evolution of the electron's bunch spectrum  is described by a transport equation,  \cite{landau44,bichsel06}:
\begin{eqnarray} \label{eq:traneq}
\frac{\partial f(x,\gamma)}{\partial x}  &=& \int_{-\infty}^\infty \left[ w(\omega,\gamma + \omega) f(x,\gamma + \omega)- \right. \\
 &&\left.  w(\omega,\gamma) f(x,\gamma )\right]\D\omega \, ,
\end{eqnarray}
with $\gamma $ being the dimensionless particle energy (Lorentz factor).

A solution to \eqref{eq:traneq} in a form of the characteristic function (Fourier transform of the distribution density), \cite{bulyak17b,bulyak18} is:
\begin{equation} \label{eq:hatf}
\hat{f} = \hat{f}_0\exp[ x(\check{w} - 1)]\; ,
\end{equation}
where $w(\omega ) $ is the recoil's spectrum, and $f_0 $ is the initial spectrum.
The parameter $x > 0$ is the ensemble average number of the recoils undergone by an electron since entering the driving force, \cite{bulyak17b,bulyak18}.

The equation \eqref{eq:hatf} may be generalized and simplified due to the model's assumption of the independence of recoils on the electron's energy, as proposed in \cite{landau44}. Instead of the beam spectrum, we consider the distribution density of losses: so called straggling function \cite{bulyak18}. The straggling function, normalized to unity, presents the loss spectrum: only the particles that have undergone at least one recoil contribute to it.

The characteristic function for the straggling function $S_x$ and the Poisson-weighted expansion are
\begin{align} \label{eq:fourstra}
\hat{S}_x  &=  \hat{w}\E^{x(\hat{w}-1)}\; , &
S_x(\epsilon )&= \sum_{n=0}^\infty \frac{\mathrm{e}^{-x} x^n}{\Gamma(n+1)} F_n(\epsilon)\; ,\\
F_n&=F_{n-1}\ast w\; , & F_0&=w\; . \nonumber
\end{align}
where $\ast $ stands for the convolution operation.
The first three moments of the straggling function---mean, variance, and skewness---read:
\begin{subequations}\label{eq:rawmom}
\begin{align}\label{eq:1raw}
\overline{\epsilon} &= (1+x)\, \overline{\omega}\; ;\\
\mathrm{Var}[\epsilon]\equiv \overline{\left(\epsilon-\overline{\epsilon}\right)^2} &=(1+x)\, \overline{\omega^2}-\overline{\omega}^2\; ;\label{eq:2raw} \\
\mathrm{Sk}[\epsilon] \equiv \overline{\left(\epsilon-\overline{\epsilon}\right)^3}&= (1+x)\, \overline{\omega^3}-3\, \overline{\omega^2}\, \overline{\omega} +2\, \overline{\omega}^3  \; . \label{eq:3raw}
\end{align}
\end{subequations}
Here $\overline{\omega},\overline{\omega^2},\overline{\omega^3}$ are the raw moments of the  recoil spectrum $w(\omega)$, $\overline{\omega^n} \equiv \int \omega^n w(\omega )\D\omega$, the `overline' sign indicates the ensemble average.

A universal solution for the straggling function \eqref{eq:fourstra} allows for accurate evaluation at the beginning of the process, $x\lesssim 1$ when the series may be limited to a few self-states $F_n$; and in the opposite limit of the large number of recoils, $x\to \infty $ when a few first moments \eqref{eq:rawmom} adequately represent the function.
The first moment---mean energy loss---always holds, since it presents the energy conservation law.

\subsection{Statistical properties for finite number of recoils}
From a practical point of view,the most interesting for the physically realizable systems is a medium number of recoils, when a particle, after entering the system, lost a small fraction of its initial energy in the moderate number of the recoils, $x\overline{\omega }\ll \gamma_0 $ with $\gamma_0$ being the initial electron energy.

To evaluate the functional dependency of the spectrum width against the average number of recoils, $\sigma = \sigma (x)$, we compare the distribution \eqref{eq:fourstra} with the L\'{e}vy $\alpha $-stable distributions, the only ones to which the sum of independent identically distributed variables is attracted, see \cite{nolan:2017}.

\begin{figure} 
   \includegraphics[width=\columnwidth]{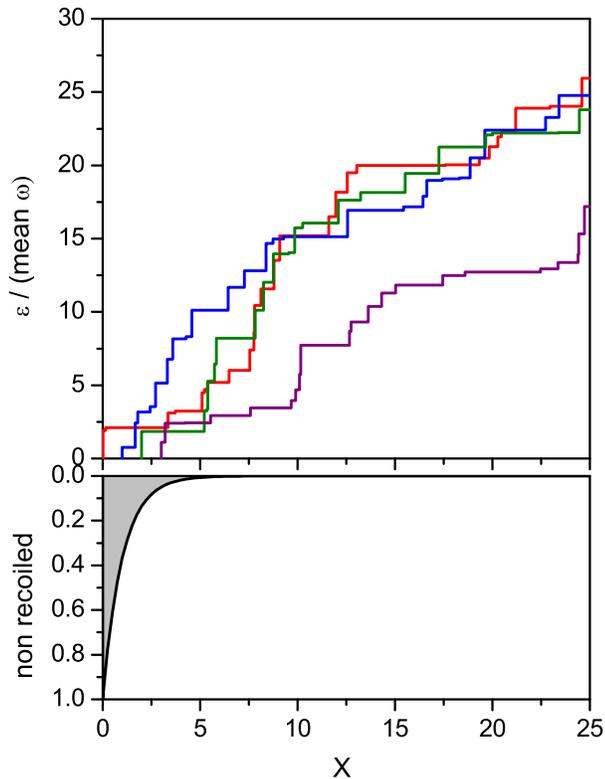}
   \caption{Four simulated trajectories for the recoils from the dipole radiation emission (top). Density of the non-recoiled particles is indicated in grey on the bottom panel.}  \label{fig:mul2}
\end{figure}

It should be emphasized, that despite the relevance of the trajectory to the $\alpha$-stable class, the bunch of such trajectories do not rigorously match the class, since individual trajectories come into stage at different $x$, as is depicted in Fig.~\ref{fig:mul2}. Nevertheless, at $x\gg 1 $ when almost all of the particles have been recoiled and the width of the distribution exceeds the width of the spectrum of recoils, the bunch of the trajectories is expected to obey a stable law.

The characteristic function of $\alpha $-stable  process, see e.g. \cite{nolan:2017}, has a general form:
\begin{equation}\label{eq:levy}
\hat{\phi}(s)=\exp\left\{ - 2\pi\I s\mu -|2\pi \sigma s|^\alpha \left[ 1-\I\beta\, \s(s)\Phi\right] \right\}\; ,
\end{equation}
with
\[
\Phi =
\begin{cases}
\tan \left(\frac{\pi \alpha}{2}\right), & \alpha\neq 1,\\
-\frac{2}{\pi}\log |s|, & \alpha = 1,
\end{cases}
\]
The parameters of the stable distribution are $\alpha\in (0,2]$ the stability parameter,  $\sigma >0 $ the scale parameter, $\beta $ the skewness parameter, and $\mu $ the location parameter.

The model under consideration allows for a reduction of range of the parameters: the stability parameter should be in the range $\alpha\in (1,2]$ due to a finite mean of the recoil spectrum; the location parameter is simply equal to the first moment of the straggling function \eqref{eq:1raw}.

Comparison of the characteristic function of straggling \eqref{eq:fourstra} with that of the stable distribution \eqref{eq:levy} leads to two important consequences: (i) the scaling parameter $\sigma $ is determined by the real part of the exponent, and (ii) the  Fourier transform of a recoil spectrum in general may not be of the power form, $\propto |s|^\alpha $ with $\alpha = \text{const}$.

Because of aiming the study at evaluation the scale parameter of distribution and taking into account the similarity theorem---the width of the distribution is inversely proportional to the width of its Fourier transform---we suggest evaluating the stability parameter at $s=s_*$ where the real part of the exponent of the characteristic function equals to unity:
\begin{equation}\label{eq:reex}
\Re [x\left( 1-\hat{w}(s_*)\right)] = 1 = \left|\pi \sigma s_*\right|^\alpha\; .
\end{equation}
Here in the square brackets of left-hand side of \eqref{eq:reex}, we intentionally omit the small term $\Re [\log \hat{w}(s_*)]\propto - s_*^2$.

From this suggestion, an expression for the stability parameter is readily derived:
\begin{equation}\label{eq:alpha}
\alpha = \left. \frac{s D_s \Re[\hat{w}]}{1-\Re[\hat{w}]}\right|_{s=s_*}\; ,
\end{equation}
where $D_s\equiv \frac{\partial }{\partial s}$ and $s_*=s_*(x) > 0 $ is the root of \eqref{eq:reex}.

Substituting the explicit expression for $\Re[\hat{w}]$,
\begin{align*}
  \left|\pi \sigma s_*\right|^\alpha &= x\int w(\omega)\left[1-\cos(-2\pi s_* \omega)\right]\D \omega \\
 &= x\int w(\omega)\left[\pi s_* \omega\right]^\alpha\D \omega =
  x m_\alpha[w]\left[\pi s_* \right]^\alpha\; ,
\end{align*}
we get a general dependence of the scale parameter---the width of the straggling distribution---on number of recoils:
\begin{equation}\label{eq:sigmax}
\sigma (x) = \left[ x m_\alpha[w] \right]^{1/\alpha}\; ,
\end{equation}
where $m_\alpha[w]$ is the raw generalized $\alpha $-moment of the recoil spectrum:
\[
m_\alpha[w]\equiv \int \omega^\alpha w(\omega)\D \omega\; .
\]

Thus, the width of the spectrum increases with the average number of the recoils as $\propto x^{1/\alpha (x)}$. The stability parameter $\alpha (x)$, in turn, increases with $x$ from unity to two.

It should be noted, that the scale parameter is equal to half-width of the distribution at $1/\E $ of the maximum. At $\alpha \to 2$ when the distribution approaches the normal (Gaussian) distribution, the scale parameter approaches the square root of Gaussian variance divided by two, $\sigma (x) \to \left[\overline{(\epsilon -\overline{\epsilon})^2}/2\right]^{1/2}$.

\section{VERIFICATION OF THE METHOD}
The two known functional limits of the considered process, $1\le \alpha \le 2$ with $\alpha =1$ being the Landau distribution and $\alpha =2 $ the Fokker-Plank (diffusive limit, the Gaussian distribution), may be considered the benchmarks of the method, see, e.g., \cite{ashrafi17}.

\subsection{The diffusion limit}
As stated in the Central Limit Theorem, the sum of independent identically distributed variables with the finite variance should approach the normal (Gaussian) distribution, which is a limiting case of the stable distributions with $\alpha =2$. Directly following from \eqref{eq:fourstra}, at $x\to \infty $, the real part of the exponent approaches Gaussian:
\[
\Re [\hat{w}-1]\approx 2\pi^2 s^2\overline{\omega^2}\; .
\]
The same result, $\alpha = 2$, directly stems from \eqref{eq:alpha} since $s_*\to 0$ when $x\to \infty $.

\subsection{The Landau distribution}
A particular case of the stable distributions, the Landau distribution function \cite{landau44} ($\alpha = 1$, $\beta = 1$), is of special importance since it has undergone extensive study and experimental validation, see \cite{vavilov57,bichsel88}.
The process of ionization losses described by the Landau distribution, agrees with the assumption of the small recoils, whose spectrum is independent of the energy of the particles. The problem is the employment of the idealized unbound recoil spectrum of $\propto \omega^{-2} $ dependence on energy. This spectrum---the Rutherford cross section---can not be normalized (it has infinite moments).

To avoid the divergence, we consider a truncated recoil spectrum, $0 < a\le \omega\le b < \infty$, then take the limits $a\to 0$, $b\to \infty $, and keep the total energy losses finite. A `physical' normalized Rutherford cross section (see \cite{linhard85,bak87}) reads:
\begin{equation} \label{eq:spenor}
w_\text{L}(\omega ) = \frac{\s (\omega -a)-\s(\omega -b)}{2 \omega ^2}\;\frac{a b}{(b-a)}\; .
\end{equation}
Its raw moments are finite:
\[
\overline{\omega }= \frac{a b}{b-a} \log \left(\frac{b}{a}\right)\; , \qquad
\overline{\omega^2}= {a b}\; .
\]

Fourier transform for this cross-section is:
\begin{align}
  \hat{w}_\text{L}(s) &=\frac{1}{(b-a) \sqrt{2 \pi }}\times  \\
  &\left\{ b \cos (a s)-a \cos (b s)+ s a b \left[\mathrm{Si}(a s)-\mathrm{Si}(b s)\right] - \right. \nonumber \\
 {} &\left. \I \left[ s a b\left( \mathrm{Ci}(a s)-\mathrm{Ci}(b s)\right) - b \sin (a s)+ a \sin (b s)\right] \right\} \nonumber
\end{align}
where  $\mathrm{Si}(z)=\int_{0}^{z}\sin (t)/t\D t$ is the integral sinus, and $\mathrm{Ci}(z)=-\int_{z}^{\infty}\cos (t)/t\D t$ is the integral cosine.

Explicitly, we have for the model:
\begin{eqnarray}\label{eq:alpLan}
\alpha_\text{L}(s;a,b) &=& 1 + \frac{a \cos (2 \pi  b s)-b \cos (2 \pi  a s)}{a b (b-a)} + \nonumber \\
&& \frac{2 \pi  s a b}{(b-a)}\left[\mathrm{Si}(2 b \pi  s)- \mathrm{Si}(2 a \pi  s)\right]\; ,
\end{eqnarray}
with $s=s_*$ being the root of \eqref{eq:reex}.

The stability parameter \eqref{eq:alpLan} has two limits: (i) when $a,b$ finite, $x\to \infty $ (accordingly $s_*\to 0$) and (ii) $x$ finite, $a\to 0, b\to \infty $ (Rutherford cross section):
\begin{subequations}\label{eq:landlimits}
\begin{align}
  \lim\limits_{s\to 0}\alpha_\text{L}(s;a,b) &= 2\,,\qquad 0<a<b<\infty \; ; \label{eq:lims0} \\
  \lim\limits_{a\to 0,b\to \infty}\alpha_\text{L}(s;a,b) &= 1\,,\qquad 0 < s\; . \label{eq:limb0}
\end{align}
\end{subequations}

Thus, the stability parameter \eqref{eq:alpLan} coincides  with the Landau distribution at the Rutherford cross section and with the Gaussian distribution at $x\to \infty$ and the finite-moments recoil spectra.

\begin{figure} 
   \includegraphics[width=\columnwidth]{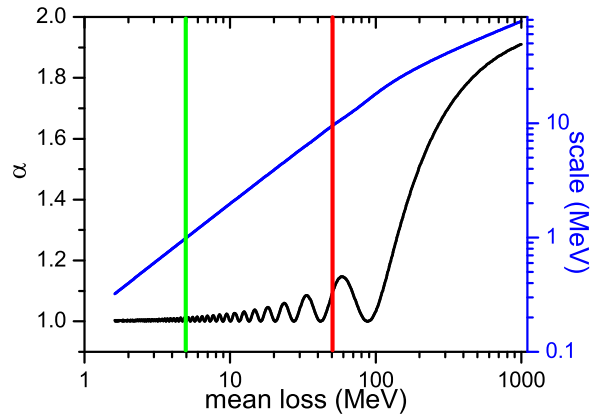}
   \caption{The stability parameter (black curve) and the scale parameter (blue) against mean energy loss. The vertical green line indicates the limit of validity of the Landau distribution (10\% loss, see \cite{payne69}), the red line indicates the physical limit: all the energy radiated out.}
   \label{fig:Las}
\end{figure}

For the physically grounded cases of ionization losses, both the Landau and the Vavilov formulas are valid. The Landau distribution evolves into the Gaussian well beyond the physical region, as is illustrated in Fig.~\ref{fig:Las} for 50\,MeV electrons traversing liquid hydrogen (in this case the ionization losses are dominant).

As it can be seen from Fig.~\ref{fig:Las}, the Landau distribution adequately describes evolution of the straggling function. The width of the distribution linearly within the range of validity linearly increases with the mean losses. (Small oscillations in the stability parameter occur because of errors in the numerical computation of the root $s_*$.)

\section{RADIATION IN PERIODIC STRUCTURE}
As an example of application of the method to a practical case, we consider evolution of the straggling function due to emission of the undulator radiation, see, e.g., \cite{howells92}.
The undulator parameter $K$, which is:
\[
K= \frac{e B\lambda_\text{u}}{2\pi m_\text{e}c}\, ,
\]
where $B$ is the magnetic field strength, $\lambda_\text{u}$ is the spatial period of the magnetic field, $e, m_\text{e}$ are the electron charge and the rest mass, resp., $c$ the speed of light.

Evolution of the straggling function for a long undulator and $K\gtrsim 1$ approximates the diffusion process \cite{agapov14}. On the other hand, for the dipole radiation $K\ll 1$, and a short undulator (or the entrance section of a long undulator), $x\lesssim 5$ this function is asymmetric and non-Gaussian \cite{bulyak15,bulyak17b}.

\begin{figure} 
   \includegraphics[width=\columnwidth]{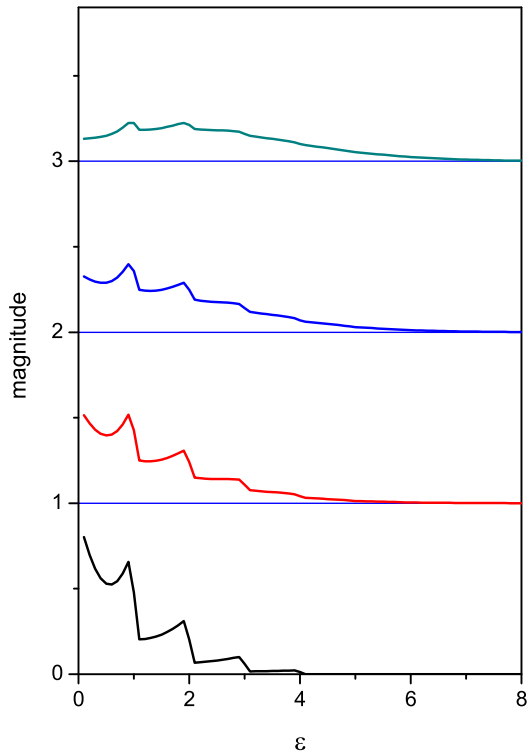}
   \caption{Straggling distribution function caused by recoils in helical undulator, $K=1$, $x=0.01,0.5,1,2$ (shifted by 0,1,2,3 from bottom to top, respectively)}
   \label{fig:spectrum4}
\end{figure}

Figure~\ref{fig:spectrum4} represents the straggling function profiles computed in accordance to \eqref{eq:fourstra} for a small average number of recoils. It shows, that the straggling function resembles the spectrum of the recoil at $x\ll 1$, then it gradually spreads out and smoothes, approximating to some degree the Landau distribution.

\begin{figure} 
   \includegraphics[width=\columnwidth]{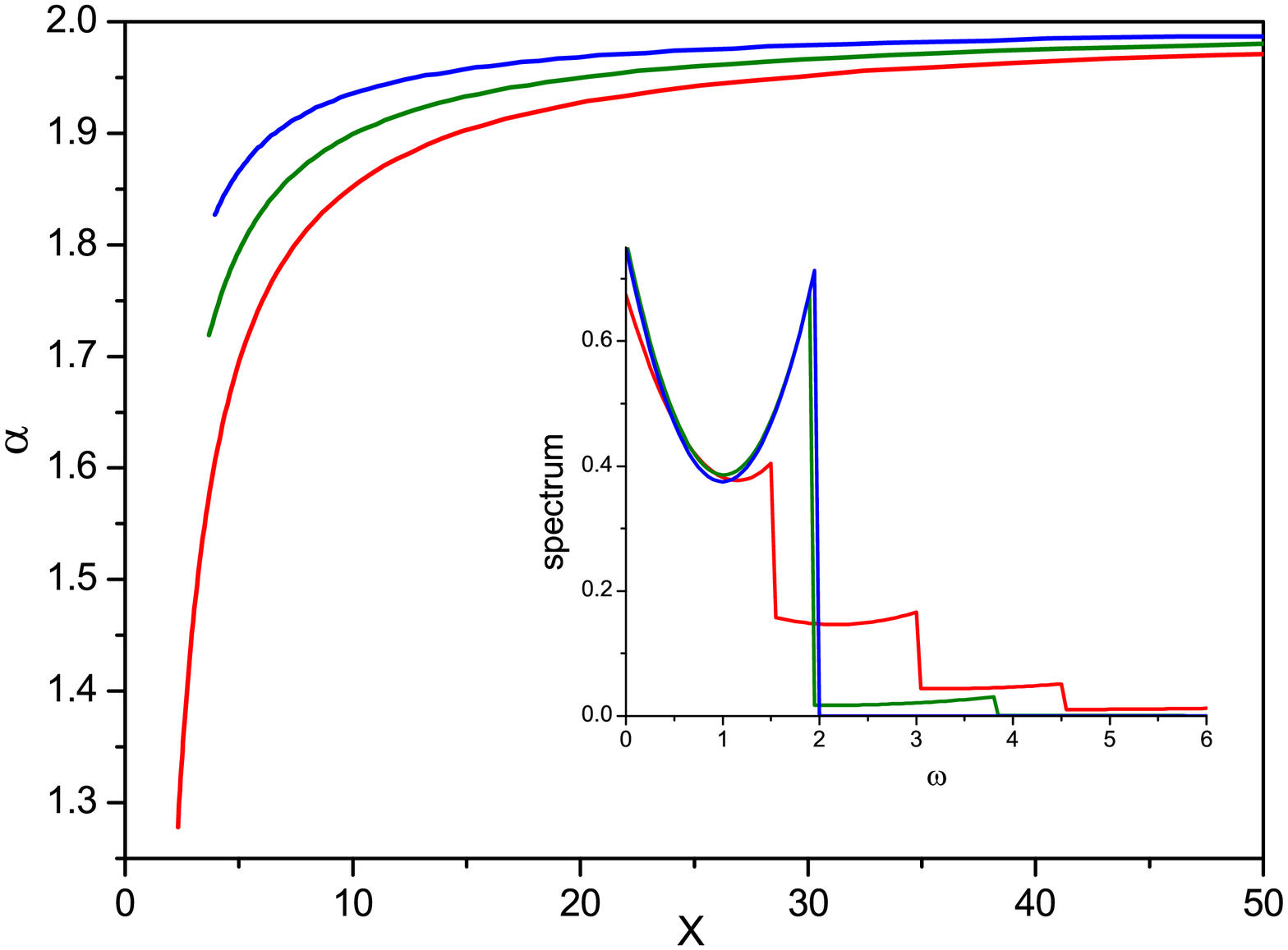}
   \caption{Stability parameter for $K =0.01$ (blue), $K =0.3$ (green), and  $K =1$ (red) vs average number of recoils. (The inset presents the corresponding recoil spectra.)}
   \label{fig:und3k}
\end{figure}

The stability parameter against the number of recoils, computed  based on \eqref{eq:alpha} for different undulator parameters $K$ is presented in Fig.~\ref{fig:und3k}. As it can be seen from the figure, the wider the recoil spectrum, the later the stability parameter approaches the diffusion limit of $\alpha = 2 $.

When the stability parameter approaches the `diffusion' value of $\alpha = 2$
(still remaining below it), the third centered moment \eqref{eq:3raw} stays positive and increases with $x$. We can derive practical information about the mode of the distribution. Making use of Pearson's skewness for a distribution close to normal, see, e.g. \cite{kenney62},
\[
\frac{\overline{\epsilon}-\epsilon_\text{mode}}{\sigma} = \frac{\mathrm{Sk}[\epsilon]}{2\sigma^3}\;,
\]
where $\epsilon_\text{mode}$ is the maximum of the distribution density,  we get:
\begin{align} \label{eq:mode}
\epsilon_\text{mode} &= \overline{\epsilon} - \frac{\mathrm{Sk}[\epsilon]}{2\sigma^2} \nonumber \\
& = (1+x)\overline{\omega}-\frac{(1+x)\overline{\omega^3}-3\overline{\omega^2} \overline{\omega}+2\overline{\omega}^3}{2\left[(1+x)\overline{\omega^2}- \overline{\omega}^2\right]}\; .
\end{align}

For a big number of the recoils, $x\to \infty $, the shift of the mode from the mean is almost independent of the number of recoils:
\[
\epsilon_\text{mode} - \overline{\epsilon} \approx - \frac{\overline{\omega^3}}{2\overline{\omega^2}}\; .
\]

The mode---position of the maximum---is shifted from the mean to smaller energy losses by the constant value, which is determined by the raw moments of the recoil spectrum.

\section{SUMMARY}
A general dependence of the distribution of energy losses by the relativistic electrons due to radiation in periodic structures or ionization losses in matter was analyzed. The straggling function---distribution density of fluctuations---is  determined solely by the ensemble-average number of recoils having undergone by the particle since entering the field (or medium in the case of ionization losses), and the spectrum of the recoil.

The straggling function was compared to the L\'{e}vy stable process as the only attractor of such processes according to the Generalized Central Limit Theorem. The  results of this consideration reveal that the width of the electron spectrum is increasing with the number of recoils according to the power law, with the power index being inverse to the stability parameter, i.e., linearly with the number of recoils at the beginning of the process,  and in proportion to the square root from the number of recoils at the diffusion limit.

Increase of the spectrum width limits the ability of the beam to generate coherent radiation in hard x-ray and gamma-ray region.

Despite of the assumed independence of the recoil spectrum on the electron's energy, a `negligible' (from the electron's point of view) change in this spectrum may play an important role in the reduction of the brightness of  sources of hard x-rays and gamma-rays, which employed relativistic electrons.

It occurs due to the fact that only a small fraction of the spectrum is used: the pin-hole fraction of the spectrum has strong dependency upon the energy spread of the electrons. The attainable width of the pin-hole collimated radiation---upper limit of it---will exceed double of the electron bunch energy spread, \cite{bulyak14a}.

\providecommand{\noopsort}[1]{}\providecommand{\singleletter}[1]{#1}

\end{document}